\documentclass[%
 aip,
jmp,%
 amsmath,amssymb,
preprint,%
showkeys
]{revtex4}

\usepackage{graphicx}% Include figure files
\usepackage{dcolumn}% Align table columns on decimal point
\usepackage{bm}% bold math
\begin{document}

\preprint{Progress in Turbulence VI}%{NORDITA-2016-***}

\title[Helicity dissipation-rate equation]{Modeling helicity dissipation-rate equation}% Force line breaks with \\

\author{Nobumitsu Yokoi}
\email{nobykoi@iis.u-tokyo.ac.jp}
\affiliation{ 
Institute of Industrial Science, University of Tokyo\\
4-6-1 Komaba, Meguro, Tokyo 153-8505, Japan
}%
\altaffiliation{Guest researcher at the Nordic Institute of Theoretical Physics (NORDITA)}%Lines break automatically or can be forced with \\

\date{\today}% It is always \today, today,
             %  but any date may be explicitly specified

\begin{abstract}
Transport equation of the dissipation rate of turbulent helicity is derived with the aid of a statistical analytical closure theory of inhomogeneous turbulence. It is shown that an assumption on the helicity scaling with an algebraic relationship between the helicity and its dissipation rate leads to the transport equation of the turbulent helicity dissipation rate without resorting to a heuristic modeling.
%
%Valid PACS numbers may be entered using the \verb+\pacs{#1}+ command.
\end{abstract}

\pacs{Valid PACS appear here}% PACS, the Physics and Astronomy
                             % Classification Scheme.
%Use showkeys class option if keyword
                              %display desired
\maketitle

%-----------------------------------------------------------------------------------
%	1. Introduction
%-----------------------------------------------------------------------------------
\section{Introduction}
\label{sec:1}
The helicity defined by $\int_V {\bf{u}} \cdot \mbox{\boldmath$\omega$}\ dV$, as well as the kinetic energy, is an inviscid invariant of the Navier--Stokes equation [$V$: fluid volume, {\bf{u}}: velocity, $\mbox{\boldmath$\omega$} (= \nabla \times {\bf{u}}$): vorticity]. Unlike the local turbulent energy density $\langle {{\bf{u}}'{}^2} \rangle /2$, the local turbulent helicity density $\langle {{\bf{u}}' \cdot \mbox{\boldmath$\omega$}'} \rangle$ is non-positive-definite and a pseudoscalar that changes its sign under the inversion or reflection (${\bf{u}}'$: velocity fluctuation, $\mbox{\boldmath$\omega$}'$: vorticity fluctuation). Since any pseudoscalar should vanish in a mirror symmetric system, a non-zero pseudoscalar represents the breakage of mirror symmetry. In non-mirror symmetric turbulence, a finite helicity density is spatially distributed to affect the local transport. The dynamic evolution of turbulent helicity is subject to the balance among the helicity production (from the large-scale inhomogeneities), its dissipation and transport rates.

	Effects of helicity (hereafter we drop ``density'') have been examined in the context of turbulent transports. In the dynamos, the turbulent helicity is directly connected to the so-called $\alpha$ effect, and plays an important role in magnetic field generation \cite{par1955,yok2013}. Also in the non-mirrosymmetric hydrodynamic turbulence such as a turbulent swirling flow, the turbulent helicity is expected to counterbalance the eddy viscosity \cite{yok1993}. The evaluation of the helicity dissipation rate is of crucial importance in determining the magnitude of effective transport.

%-----------------------------------------------------------------------------------
%	2. Helicity in inhomogeneous turbulence
%-----------------------------------------------------------------------------------
\section{Helicity in inhomogeneous turbulence}
\label{sec:2}
The turbulent helicity $H \equiv \langle {{\bf{u}}' \cdot  \mbox{\boldmath$\omega$}'} \rangle$ obeys an exact transport equation:
\begin{equation}
	{DH}/{Dt} 
	\equiv \left( {{\partial}/{\partial t} + {\bf{U}}\cdot \nabla} \right)H 
	= P_H 
	- \epsilon_H 
	+ \nabla \cdot T_H
	\label{eq:H_eq}%(1)
\end{equation}
(${\bf{U}}$: mean velocity). Here, $P_H$ and $T_H$ are the production and transport rates of $H$, whose expressions are suppressed. The helicity dissipation rate $\epsilon_H$ is defined by
\begin{equation}
	\epsilon_H \equiv 2\nu \left\langle {
		{\frac{\partial u'{}_b}{\partial x_a}}
		{\frac{\partial \omega'{}_b}{\partial x_a}}
	} \right\rangle.
	\label{eq:hel_diss_def}%(2)
\end{equation}
Evaluation of $\epsilon_H$ is of crucial importance to estimate the turbulent helicity evolution.

	In order to derive the dynamic equation of $\epsilon_H$, we have to express $H$ in inhomogeneous turbulence. We follow the formal procedure of the two-scale direct-interaction approximation (TSDIA) \cite{yos1984,yok1993,yok2013}, a combination of the multiple-scale analysis with a propagator renormalization closure theory of turbulence. In the TSDIA framework, the lowest-order velocity field is equivalent to the homogeneous isotropic turbulence, and the effects of the mean-field inhomogeneities, rotation, etc.\  are taken into account in a perturbation manner in the first- and higher-order velocity fields. If we introduce the Green's function of the lowest-order velocity field, $G'_{{\rm{B}}\alpha\beta}({{\bf{k}};\tau,\tau'})$, the first-order velocity field ${\bf{u}}'_1$ can be expressed in terms of this Green's function. For the lowest order velocity field ${\bf{u}}'_0$ and the Green's function, we assume the generic form for the homogeneous isotropic turbulence as
\begin{equation}
	\frac{
	{\langle {
		u'_{0\alpha}  \left( {{\bf{k}}{\rm{;}}\tau } \right)
		u'_{0\beta}  \left( {{\bf{k'}}{\rm{;}}\tau'} \right)
	} \rangle}}
	{{\delta \left( {{\bf{k}} + {\bf{k'}}} \right)}
	} 
	= D_{\alpha\beta} \left( {\bf{k}} \right)
	Q_{\rm{B}} \left( {k{\rm{;}}\tau ,\tau '} \right) 
	+ \frac{i}{2} \frac{{k_a}}{{k^2}} \epsilon_{\alpha\beta a} 
	H_{\rm{B}} \left( {k{\rm{;}}\tau,\tau'} \right),
	\label{eq:homo_iso_u0}%(3)
\end{equation}
\begin{equation}
	\langle {
	G'_{{\rm{B}}\alpha\beta} \left( {{\bf{k}};\tau,\tau'} \right)
	} \rangle
	= D_{\alpha\beta} \left( {\bf{k}} \right)
	G_{\rm{B}} \left( {k;\tau,\tau'} \right),
	\label{eq:homo_iso_G}%(4)
\end{equation}
where $D_{\alpha\beta}({\bf{k}}) = \delta_{\alpha\beta} - k_\alpha k_\beta / k^2$ is the projection operator. Here, $Q_{\rm{B}}$ and $H_{\rm{B}}$ are the spectral functions of the kinetic energy and helicity of the lowest-order fields, respectively. The second term in Eq.~(\ref{eq:homo_iso_u0}) represents the non-mirror symmetry of turbulence with $H$ being a pseudoscalar.

	The turbulent helicity is expanded as
\begin{equation}
	H({\bf{x}};t) = \left\langle {
		{\bf{u}}' \cdot {\mbox{\boldmath$\omega$}'}
	} \right\rangle
	= \left\langle {
		{\bf{u}}'_{\rm{0}} \cdot \mbox{\boldmath$\omega$}'_{\rm{0}}
	} \right\rangle
	+ \left\langle {
		{\bf{u}}'_{\rm{0}} \cdot \mbox{\boldmath$\omega$}'_{1}
	} \right\rangle
	+ \left\langle {
		{\bf{u}}'_1 \cdot \mbox{\boldmath$\omega$}'_{\rm{0}}
	} \right\rangle
	+ \cdots.
	\label{eq:H_perturb_exp}%(5)
\end{equation}
Substituting ${\bf{u}}'_0$ and ${\bf{u}}'_1$ ($\mbox{\boldmath$\omega$}'_n = \nabla \times {\bf{u}}'_n$ with $n=0,1$) into Eq.~(\ref{eq:H_perturb_exp}), with a renormalization procedure ($Q_{\rm{B}} \to Q$, $H_{\rm{B}} \to H$, $G_{\rm{B}} \to G$), we obtain 
\begin{equation}
	H({\bf{x}};t)
	= I_0 \{H\}
	- \frac{1}{3} I_0 \left\{ {G, \frac{DH}{Dt}} \right\}
	+ \frac{8}{27} \left( {
		\mbox{\boldmath$\Omega$} 
		+ 2 \mbox{\boldmath$\omega$}_{\rm{F}}
	} \right) \cdot I_{0}\left\{ {G, \nabla Q} \right\},
	\label{eq:H_expression}%(6)
\end{equation}
where $\mbox{\boldmath$\Omega$} (= \nabla \times {\bf{U}})$ is the mean vorticity, $\mbox{\boldmath$\omega$}_{\rm{F}}$ the angular velocity, and the abbreviated forms of integral are defined by 
\begin{equation*}
	I_n \left\{ {A} \right\}
	= \int k^{2n} {A(k,{\bf{x}}; \tau,\tau,t)} d{\bf{k}} 
\end{equation*}
and 
\begin{equation*}
	I_n \left\{ {A, B} \right\}
	= \int k^{2n}\! d{\bf{k}} \int_{-\infty}^\tau d\tau_1
	\nonumber
	A(k,{\bf{x}}; \tau,\tau_1,t) B(k,{\bf{x}}; \tau,\tau_1,t).
\end{equation*}

	In order to evaluate integrals in Eq.~(\ref{eq:H_expression}), we assume the propagators (correlation and response functions) in the inertial range such as 
\begin{equation*}
	Q(k,{\bf{x}};\tau,\tau',t)
	= \sigma_K(k,{\bf{x}};t) \times
	\exp\left[ {-\omega_K(k,{\bf{x}};t)|\tau - \tau'|} \right],
\end{equation*}
\begin{equation*}
	H(k,{\bf{x}};\tau,\tau',t)
	= \sigma_H(k,{\bf{x}};t)
	\exp\left[ {-\omega_H(k,{\bf{x}};t)|\tau - \tau'|} \right],
	\label{eq:H_assump}%(12)
\end{equation*}
\begin{equation*}
	G(k,{\bf{x}};\tau,\tau',t)
	= \theta(\tau - \tau')
	\exp\left[ {-\omega(k,{\bf{x}};t)(\tau - \tau')} \right],
	\label{eq:G_assump}%(13)
\end{equation*}
where the spectra in the inertial range are assumed as
\begin{equation}
	\sigma_K(k,{\bf{x}};t) 
	= \sigma_{K0} \epsilon^{2/3} k^{-11/3},\;\;
	\sigma_H(k,{\bf{x}};t) 
	= \sigma_{H0} \epsilon^{-1/3} \epsilon_H({\bf{x}};t) k^{-11/3},
	\label{eq:en_hel_spec}%(7)
\end{equation}
with the time scales
\begin{equation*}
	\omega_{K}(k,{\bf{x}};t)
	= \omega_{K0} \varepsilon^{1/3} k^{2/3}
	= \tau_{K}^{-1},
\end{equation*}
\begin{equation*}
	\omega_{H}(k,{\bf{x}};t)
	= \omega_{H0} \varepsilon_H^{1/3} k^{2/3}
	= \tau_{H}^{-1},
\end{equation*}
\begin{equation*}
	\omega(k,{\bf{x}};t)
	= \omega_0 \varepsilon^{1/3} k^{2/3}
	= \tau^{-1}.
\end{equation*}
The helicity spectrum $\sigma_H$ in Eq.~(\ref{eq:en_hel_spec}) arises from the assumption that the spectrum of the helicity is determined by the scale ($k$), energy and helicity transfer rates ($\epsilon$ and $\epsilon_H$). This has been confirmed by DNSs \cite{bae2008,les2011}.

	Using Eq.~(\ref{eq:en_hel_spec}), $H$ [Eq.~(\ref{eq:H_expression})] can be estimated up to the first-order as
\begin{eqnarray}
	H({\bf{x}};t)
	&=& 3 \cdot (2\pi)^{1/3} \sigma_{H0}
		\epsilon^{-1/3} \epsilon_H \ell_{\rm{C}}^{2/3}
	+ \frac{1}{6 \cdot (2 \pi)^{1/3}} 
	\frac{\sigma_{H0}}{\omega_0 + \omega_{H0}}
	\left( {\ell_{\rm{C}} \epsilon^{-1/3}} \right)^2 \epsilon_H
	\nonumber\\
	&&\hspace{-32pt} \times \left[ {
		\left( {1 + \frac{\omega_{H0}}{\omega_0 + \omega_{H0}}} \right)
		\frac{1}{\epsilon} \frac{D\epsilon}{Dt}
		- \frac{1}{\epsilon_H} \frac{D\epsilon_H}{Dt}
		- \left( {11 + 2 \frac{\omega_{H0}}{\omega_0 + \omega_{H0}} } \right) 
			\frac{1}{\ell_{\rm{C}}} \frac{D\ell_{\rm{C}}}{Dt}
	} \right],
	\label{eq:H_exp_first}%(8)
\end{eqnarray}
where $\ell_{\rm{C}}$ is the size of the largest energy-containing eddies.

%-----------------------------------------------------------------------------------
%	3. Modeling the helicity dissipation-rate equation
%-----------------------------------------------------------------------------------
\section{Modeling the helicity dissipation-rate equation}
	In constructing a system of model equations, we can choose any three of four turbulence statistical quantities $(H, \epsilon, \epsilon_H, \ell_{\rm{C}})$. In order that any choice among the four quantities should be equivalent ({\it model transferability}), some algebraic relation should be held among them \cite{yos1987,yok2011}. 
	
	 We solve Eq.~(\ref{eq:H_exp_first}) concerning $\ell_{\rm{C}}$ in a perturbation manner. Up to the lowest-order analysis, we have
\begin{equation}
	\ell_{\rm{C}} 
	= 3^{-3/2} (2\pi)^{-1/2} \sigma_{H0}^{-3/2}
	\epsilon^{1/2} \epsilon_H^{-3/2} H^{3/2},
	\label{eq:ell_C_algeb}%(9)
\end{equation}
or equivalently, 
\begin{equation}
	\epsilon_H 
	= C_H {H}/{\tau},\;\;
	\tau
	= \ell_{\rm{C}}^{2/3} \epsilon^{-1/3},\;\;
	C_H = {1}/{[3 (2\pi)^{1/3} \sigma_{H0}]}.
	\label{eq:eps_H_algeb}%(10)
\end{equation}
As the lowest-order analysis of the turbulent helicity expression, we obtained an algebraic model for the turbulent dissipation rate as the first of Eq.~(\ref{eq:eps_H_algeb}) with the usual eddy turn-over time scale [the second of Eq.~(\ref{eq:eps_H_algeb})]. This corresponds to the estimate of the turbulent helicity dissipation rate in homogeneous isotropic turbulence.

	If we proceed to the first-order analysis, under the requirement of model transferability, the second term of Eq.~(\ref{eq:H_exp_first}) should be balanced in itself. Using Eq.~(\ref{eq:ell_C_algeb}), we change expression given in Eq.~(\ref{eq:H_exp_first}) based on $\varepsilon$, $\varepsilon_H$, and $\ell_{\rm{C}}$ into the one based on $\varepsilon$, $\varepsilon_H$, and $H$. As this result, we have
\begin{equation}
	\frac{D\varepsilon_H}{Dt}
	= C_{H1} (\omega_0, \omega_{H0}) 
	\frac{\varepsilon_H}{\varepsilon} \frac{D\varepsilon}{Dt}
	+ C_{H2} (\omega_0, \omega_{H0})
	\frac{\varepsilon_H}{H} \frac{DH}{Dt},
	\label{eq:eps_H_CH1_CH2}%(11)
\end{equation}
where $C_{H1}(\omega_0, \omega_{H0})$ and $C_{H2}(\omega_0, \omega_{H0})$ are coefficients determined by the time scales of turbulence. If we assume $\tau \simeq \tau_K \simeq \tau_H$ ($\omega_0 \simeq \omega_{K0} \simeq \omega_{H0}$), we have $C_{H1}(\omega_0, \omega_{H0}) \simeq 0.26$ and $C_{H2}(\omega_0, \omega_{H0}) \simeq 1.1$. Finally, we obtain 
\begin{equation}
	\frac{D\varepsilon_H}{Dt}
	= C_{\varepsilon_H 1} \frac{\varepsilon_H}{K} P_K
	- C_{\varepsilon_H 2} \frac{\varepsilon_H}{K} \varepsilon
	+ C_{\varepsilon_H 3} \frac{\varepsilon_H}{H} P_H
	- C_{\varepsilon_H 4} \frac{\varepsilon_H}{H} \varepsilon_H, 
	\label{eq:eps_H_eq}%(12)
\end{equation}
where the model constants are theoretically estimated as
\begin{equation}
	C_{\varepsilon_H 1} = 0.36,\;
	C_{\varepsilon_H 2} = 0.49,\;
	C_{\varepsilon_H 3} = C_{\varepsilon_H 4} = 1.1.
	\label{eq:model_consts}%(13)
\end{equation}

\section{Conclusion} 
From the lowest-order analysis, the helicity dissipation rate is estimated by an algebraic form [Eq.~(\ref{eq:eps_H_algeb})]. Up to the first-order analysis, the $\epsilon_H$ equation is derived as Eq.~(\ref{eq:eps_H_eq}) with the theoretically-determined model constants. Reflecting the spectral form [Eq.(\ref{eq:en_hel_spec})], it depends on both the energy and helicity equations.

\begin{acknowledgments}
Basic calculations of this work were performed during NY's stay at the Rudolf Peierls Centre for Theoretical Physics, The University of Oxford (January 2013) and at the Consorzio RFX in Padova (February 2013) as a visiting researcher. Part of this work is supported by the Japan Society for the Promotion of Science (JSPS) Core-to-Core Program (No.~22001) Institutional Program for Young Researcher Overseas Visits and also by the JSPS Grants-in-Aid for Scientific Research (No.~24540228).
\end{acknowledgments}

\nocite{*}
%\bibliography{n_yokoi_reference}% Produces the bibliography via BibTeX.

\end{document}